\documentclass[iop,apj,twocolumn]{emulateapj}
\usepackage{graphicx}
\usepackage{amssymb}
\usepackage{amsmath}
\usepackage{epstopdf}

\usepackage{natbib}
\submitted{To appear in Astrophysical Journal: Submitted February 2014, Revised November 2014, Accepted December 2014}

\newcommand{\kms}{km~s$^{-1}$}

\def\icarus{\ref@jnl{Icarus}}           

\begin{document}

\title{\sc  WISEP J004701.06+680352.1: An intermediate surface gravity, dusty brown dwarf in the AB Dor Moving Group}

\author{John E.\ Gizis,\altaffilmark{1,2}
Katelyn N.\ Allers, \altaffilmark{3,4}
Michael C.\ Liu,\altaffilmark{4,5}
Hugh C.\ Harris,\altaffilmark{6}
Jacqueline K.\ Faherty,\altaffilmark{7,8}
Adam J.\ Burgasser,\altaffilmark{9}
J. Davy Kirkpatrick\altaffilmark{10}
}

\altaffiltext{1}{Department of Physics and Astronomy, University of Delaware, Newark, DE 19716, USA}
\altaffiltext{2}{Visiting Astronomer, Kitt Peak National Observatory, National Optical Astronomy Observatory, which is operated by the Association of Universities for Research in Astronomy (AURA) under cooperative agreement with the National Science Foundation.}
\altaffiltext{3}{Department of Physics and Astronomy, Bucknell University, Lewisburg, PA 17837, US}
\altaffiltext{4}{Visiting Astronomer at the Infrared Telescope Facility, which is operated by the University of Hawaii under Cooperative Agreement no. NNX-08AE38A with the National Aeronautics and Space Administration, Science Mission Directorate, Planetary Astronomy Program.}
\altaffiltext{5}{Institute for Astronomy, University of Hawaii, 2680 Woodlawn Drive, Honolulu HI 96822}
\altaffiltext{6}{US Naval Observatory, Flagstaff Station, 10391 West Naval Observatory Road, Flagstaff, AZ 86001}
\altaffiltext{7}{Department of Terrestrial Magnetism, Carnegie Institution of Washington 5241 Broad Branch Road NW, Washington, DC 20015, USA}
\altaffiltext{8}{Department of Astrophysics, American Museum of Natural History, Central Park West at 79th Street, New York, NY 10034, USA}
\altaffiltext{9}{Center for Astrophysics and Space Science, University of California San Diego, La Jolla, CA 92093, USA}
\altaffiltext{10}{Infrared Processing and Analysis Center, MS 100-22, California Institute of Technology, Pasadena, CA 91125, USA}

\begin{abstract}
We present spectroscopy, astrometry, and photometry of the brown dwarf WISEP J004701.06+680352.1 (W0047+68), an unusually red field L dwarf at a distance of $12.2 \pm 0.4$ parsecs. The three-dimensional space motion identifies it as a member of the AB Dor Moving Group, an identification supported by our classification of W0047+68 as intermediate surface gravity (INT-G) using the Allers \& Liu (2013) near-infrared classification system. This moving group membership implies near-solar metallicity, age $\sim 100-125$ Myr, $M \approx 0.018~M_\odot$, and $\log g \approx 4.5$; the thick condensate clouds needed to explain the infrared spectrum are therefore a result of the lower surface gravity than ordinary field brown dwarfs. From the observed luminosity and evolutionary model radius, we find $T_{eff} \approx 1300 $K, a temperature normally associated with early T dwarfs. Thick clouds are also used to explain the spectral properties of directly imaged giant planets, and we discuss the successes and challenges for such substellar models in matching the observed optical and infrared spectra. W0047+68 shows that cloud thickness is more sensitive to intermediate surface gravity than in most models. We also present a trigonometric parallax of the dusty L6 dwarf 2MASS J21481628+4003593. It lies at $8.060 \pm 0.036$ parsecs; its astrometry is consistent with the view that it is older and metal-rich.   
\end{abstract}

\keywords{brown dwarfs ---  infrared: stars --- stars: individual: WISEP J004701.06+680352.1 --- stars: individual: 2MASS J21481628+4003593 }

\section{Introduction\label{intro}}

One of the key challenges in substellar astronomy is to use observations of colors, absolute magnitudes, and spectra to deduce the fundamental physical parameters of brown dwarfs and gas giant planets. Determining the luminosity, effective temperature, surface gravity, composition, mass and radius would allow many astrophysical applications, including testing formation scenarios for ``planetary mass objects," but is difficult in part due to the importance of condensate cloud formation at substellar atmospheric temperatures. The vast majority of the nearly two thousand known field brown dwarfs can be described by a well defined, one-dimensional spectral sequence of late-M, L, T and Y dwarfs, with a dramatic change in cloud properties at the L/T transition around 1400K (see the review of \citealt{2005ARA&A..43..195K}). It is now clear, however, that many of the directly imaged planetary mass companions have significant differences from ordinary field brown dwarfs. The companions \object[NAME 2M1207b]{2M1207b} \citep{2004A&A...425L..29C,2010A&A...517A..76P} and \object{HR 8799}b, c, d, and e \citep{Marois:2008ul,2010Natur.468.1080M}, 
are considerably redder than field brown dwarfs of similar temperature or luminosity. The planet \object[beta Pic b]{$\beta$ Pic b} has apparently normal near-infrared colors for a mid-L dwarf but very red mid-infrared colors \citep{2013ApJ...776...15C}, and the L' magnitudes of the HR 8799 planets are also brighter than expected \citep{Skemer:2012qy}. Spectra of the HR 8799 planets show considerable diversity which may require non-equilibrium mixing, different cloud properties, and perhaps enhanced metals or non-solar abundances \citep{Bowler:2010lp,Barman:2011fk,Marley:2012lr,2013Sci...339.1398K,Oppenheimer:2013fj}, but distinguishing between these possibilities is difficult.  

Rarer classes of field brown dwarfs, especially the unusually dusty ones, offer a way forward, as they can be more accessible to observations than planets.  A number of systematic studies of young ($<150$ Myr) late-M and early-L type brown dwarfs \citep{2008ApJ...689.1295K,2009AJ....137.3345C,Patience:2012fk,Faherty:2013lr,Faherty:2013qy,2013ApJ...772...79A,2014A&A...564A..55M} have demonstrated their red colors and other spectral peculiarities, which can be ascribed to low-surface-gravity dusty atmospheres. Unfortunately, even cooler brown dwarfs are necessarily fainter and more difficult to discover. \citet{Gizis:2012fk} [hereafter Paper I] noted there were then only five known field L dwarfs with $J-K > 2.3$, and none had trigonometric parallaxes. The field is developing rapidly with new sky surveys, and the two currently reddest ($J-K=2.8$) field L dwarf known are PSO J318.5338-22.8603 \citep{2013ApJ...777L..20L} and ULAS J222711-004547 \citep{2014MNRAS.439..372M}.\footnote{All names are abbreviated after first use: PSO J318-22 for  PSO J318.5338-22.8603, W0047+68 for WISEP J004701.06+680352, etc.} PSO J318-22's trigonometric parallax, tangential velocity and unusual spectrum suggests it is a planetary-mass member of the \object[NAME beta Pic moving group]{$\beta$ Pic Moving Group} \citep{2001ApJ...562L..87Z}, and in many ways is similar to 2M1207b. It has the faintest $J$-band absolute magnitude (14.8) of any L dwarf other than 2M1207b (16.13, \citealt{2007ApJ...669L..45G}). Besides PSO J318-22, the best studied of the extremely red L dwarfs are \object{2MASS J21481628+4003593} \citep{Looper:2008lr}, which is thought to be old ($\gtrsim 200$ Myr),  \object{2MASSW J2244316+204343} \citep{dahn}, which has been proposed to be young \citep{2008ApJ...689.1295K}, and \object{2MASS J035523.37+113343.7}, a dusty, red L-type dwarf with low surface gravity \citep{2009AJ....137.3345C, Faherty:2013qy,Liu:2013lr}. 

\object{WISEP J004701.06+680352} (Paper I) was the reddest ($J-K_s = 2.55 \pm 0.08$ ) known L dwarf at the time of discovery.  
Although its low-resolution near-infrared spectrum resembles older L dwarfs (see Paper I), \citet{Thompson:2013fk} have suggested it is young and low surface gravity on the basis of its moderate-resolution spectrum, and \citet{2014ApJ...783..121G} argue it and 2M2244+20 are probable members of the AB Dor Moving Group (ABDMG). In this paper, we present new observations and analysis of the brown dwarf W0047+68, including optical and near-infrared spectra, photometry, and a preliminary trigonometric parallax, as well as a parallax for 2M2148+40. We argue that W0047+68 is a {\it bone fide} member of the AB Dor Moving Group (ABDMG), and confront  theoretical dusty models with our new observational constraints. We find that 2M2148+40's space motion is consistent with the view that it is older and metal-rich.  

\section{Data and Observations\label{sec-data}}

\subsection{Near-Infrared Spectra\label{sec:nearir}}

\begin{figure*}
\includegraphics*{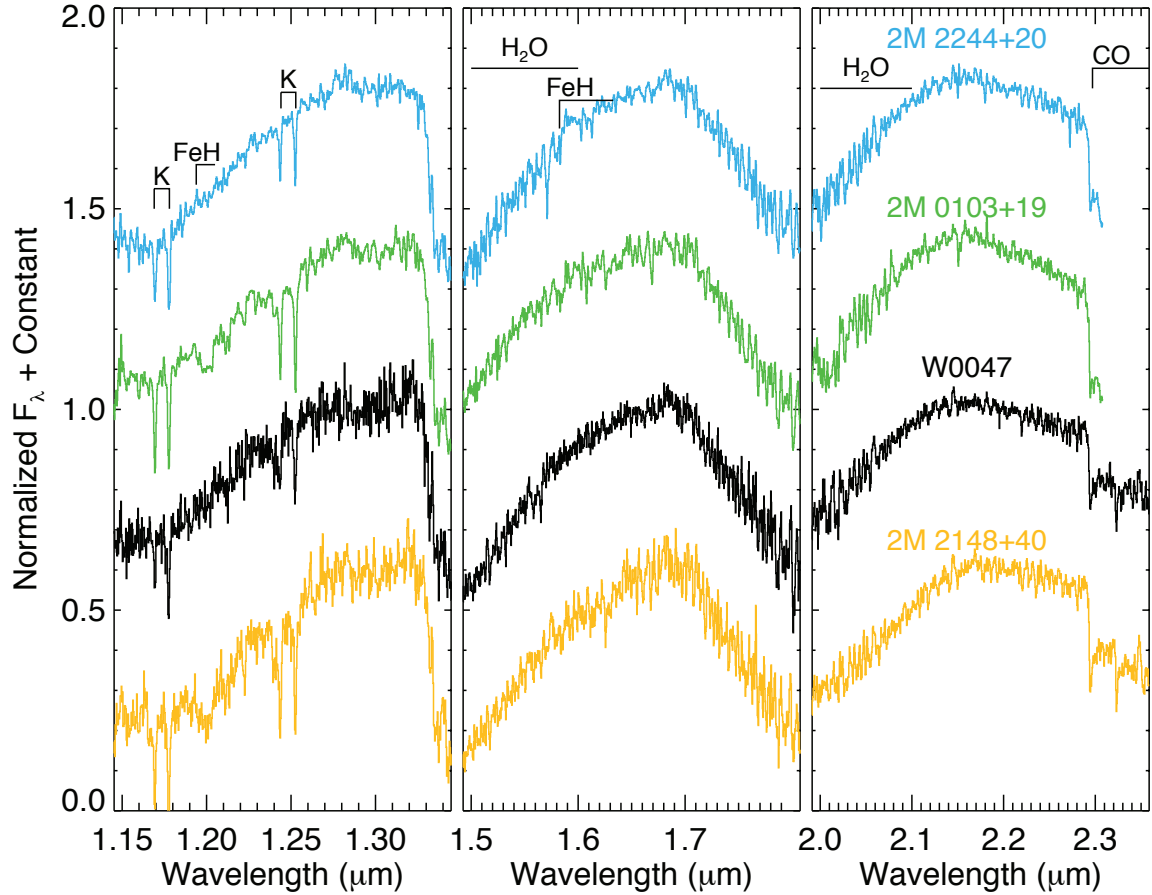}
\caption{Moderate resolution near-infrared IRTF spectrum of W0047+68 compared to other unusually red L dwarfs from \citet{2013ApJ...772...79A}. \label{fig-irallers}}
\end{figure*}

A moderate-resolution near-infrared spectrum covering the range 0.8 to 2.4 \micron~was obtained using the IRTF SpeX spectrograph \citep{2003PASP..115..362R} in cross-dispersed mode (hereinafter SXD) on UT Date 26 September 2012. The resolution is 1200. The data were reduced using the facility pipeline Spextool \citep{2004PASP..116..362C}.  Telluric features were corrected using a nearby A star using the method of \citet{2003PASP..115..389V}.  Spectra with the same observing setup were used by \citet{2013ApJ...772...79A} to develop a spectroscopic classification system that is sensitive to surface gravity: The key indicators include the \ion{K}{1} lines in the J-band 
and the FeH bands in J and H-band.  In Figure~\ref{fig-irallers}, we compare our spectrum to other unusually red L dwarfs 2M2244+20, 2MASS J01033203+1935361 \citep{2000AJ....120..447K,Faherty:2012qy}, and 2M2148+40. W0047+68 is similar to 2M2244+20, but considerably lower surface gravity than 2M2148+40.  Applying the \citet{2013ApJ...772...79A} system, we classify W0047+68 as L7 INT-G (Table~\ref{tab1}).

\begin{figure*}
\plottwo{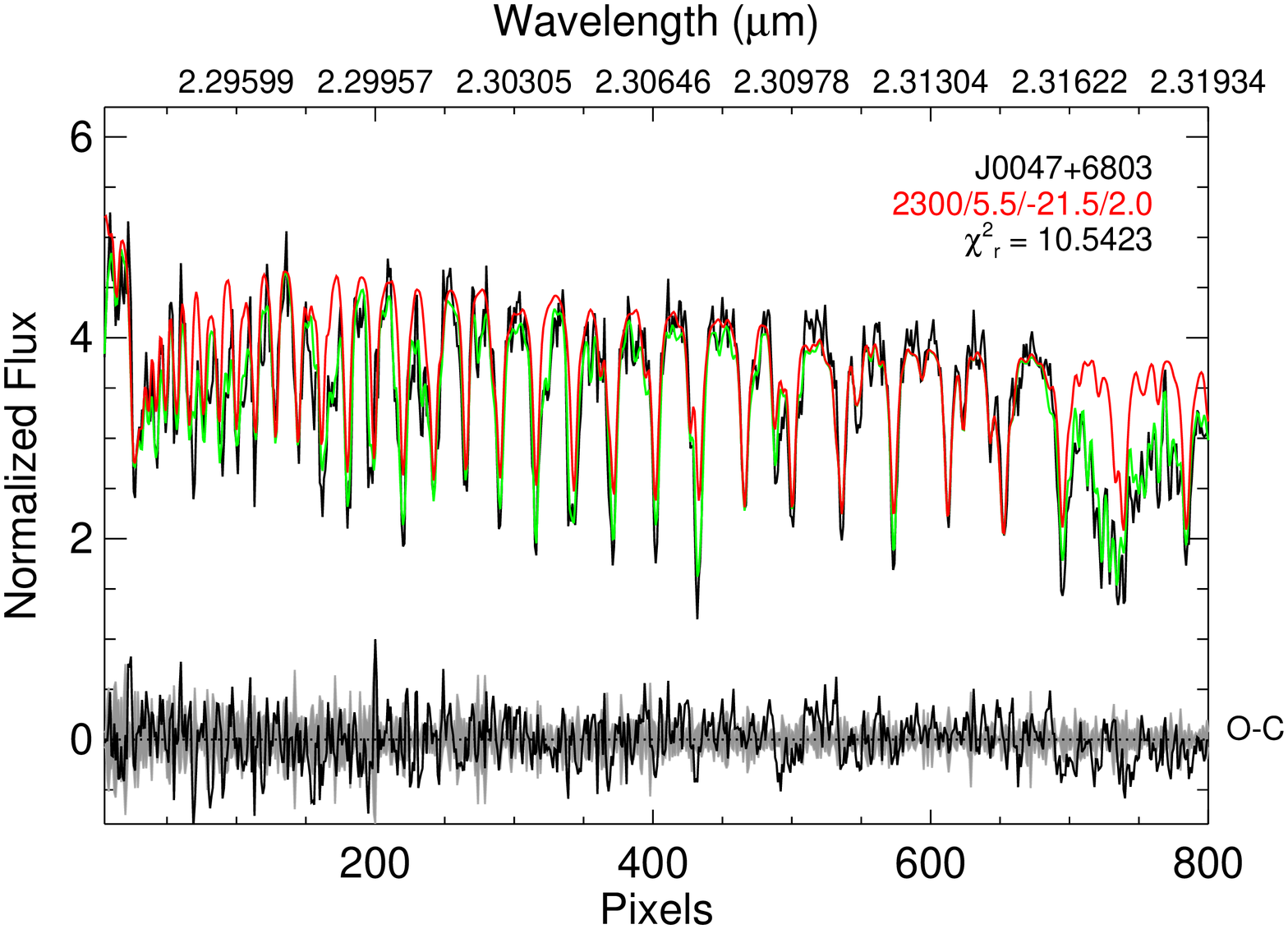}{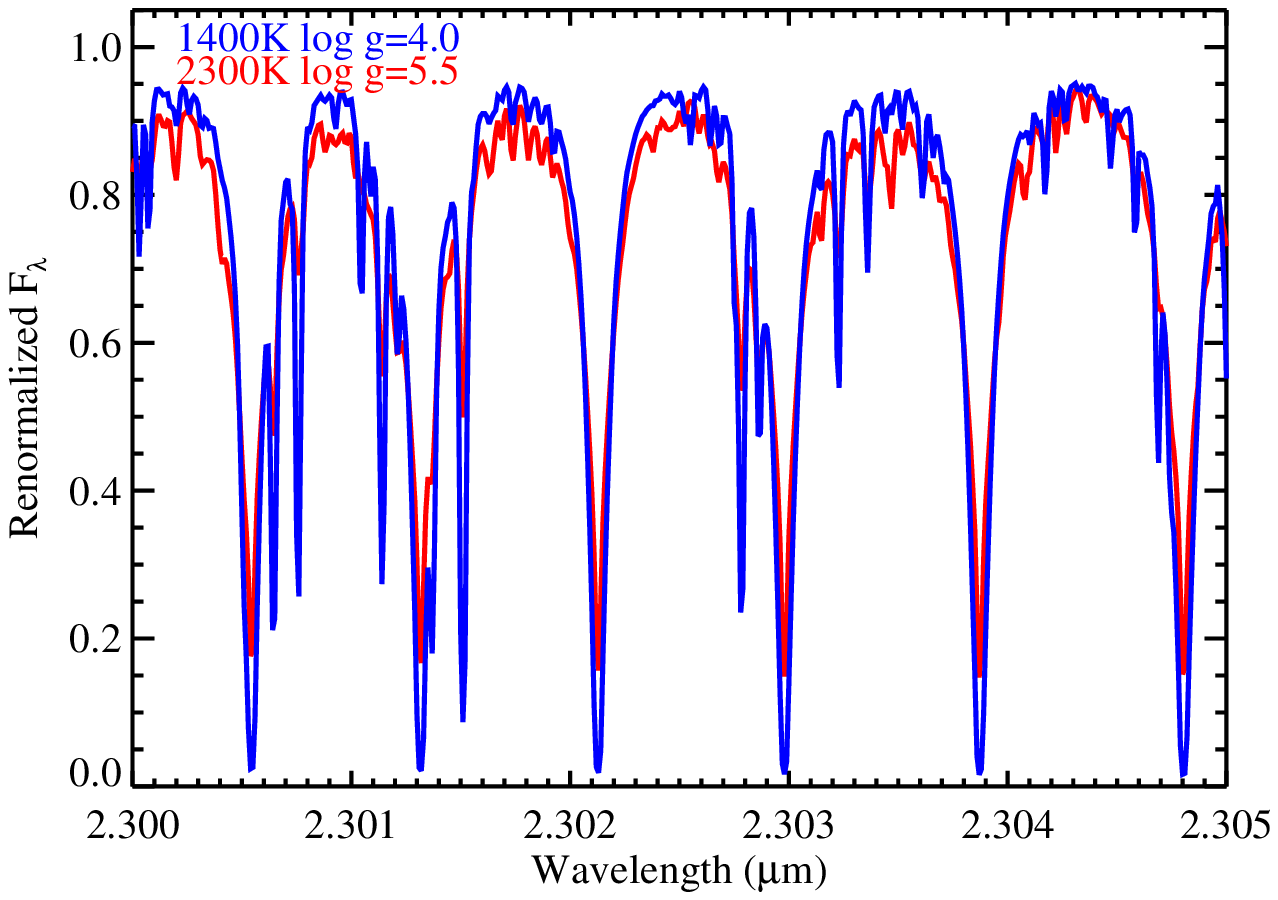}
\caption{Analysis of our Keck/NIRSPEC spectrum of W0047+68 (black line).  Left: The best-fit model (red line; T$_{eff}$ = 2300~K and $\log~g$ = 5.5 cgs) has been broadened by $v~\sin~i$ = 2~\kms~and shifted by a heliocentric radial velocity of $-$21.5~\kms.  The green line shows the combination of model and telluric absorption (from \citealt{Livingston:1991fj}) which best fits the data.  Residuals are shown at bottom (black line) and are comparable to the noise ($\pm$1$\sigma$ indicated by grey region around zero flux). Right: A comparison of two BT-Settl \citep{Allard:2011uq} models, T$_{eff}$ = 2300K with $\log~g = 5.5$ and T$_{eff}$ = 1400K with $\log~g = 4.0$, illustrating that high resolution spectra in this wavelength region can be used for velocity information but do not reliably constrain temperature.}
\label{fig:nirspec}
\end{figure*}

High resolution near-infrared spectra of W0047+68 were obtained with Keck/NIRSPEC \citep{1998SPIE.3354..566M, McLean:2000lr} on UT Date 17 Sep 2013 to determine W0047+68's radial velocity and projected rotational velocity ($v \sin i$).  Data were obtained using the N7 filter and 0$\farcs$432$\times$12$\arcsec$ slit (3-pixel dispersion), providing 1.99-2.39~$\micron$ spectroscopy in seven orders at a measured resolution of 25000 ($\Delta{v}$ = 12~\kms).  Two dithered exposures of 1200~s were obtained at an airmass of 1.50, followed by observations of a nearby A star.  Data in order~33, spanning 2.293--2.319~$\micron$, were extracted and analyzed due to the strong CO bands present in this band.  We used a combination of REDSPEC\footnote{See \url{http://www2.keck.hawaii.edu/inst/nirspec/redspec/index.html}.} and custom IDL routines that perform a forward-modeling, Monte Carlo Markov Chain analysis of the extracted spectrum using solar-metallicity BT-Settl models \citep{Allard:2011uq} and a telluric absorption spectrum from \citet{Livingston:1991fj}; this analysis will be described in detail in a forthcoming paper (Burgasser et al., in prep.; see \citealt{Blake:2010gf} for a similar algorithm). Figure~\ref{fig:nirspec} displays our extracted spectrum and the best fit model. 
The individual best fit model has $v_{rad} = -21.5$~\kms~ and $v \sin i = 2$~\kms, but marginalizing over the fit parameters including other temperatures and gravities, we find $v \sin i$ = 4.3$\pm$2.2~\kms~ and a heliocentric radial velocity of $-$20.0$\pm$1.4~\kms, which we adopt as our best estimates for W0047+68.
Our modeling requires higher resolution than is publicly available for the full BT-Settl grid; the subset of the grid that we use has $T_{eff} > 2000$K, hotter than we believe W0047+68 can possibly be. However, these CO lines are not very sensitive to temperature.  We show in Figure~\ref{fig:nirspec}b, a comparison of the $T_{eff}=1400$K, $\log g =4.0$ model (which we discuss later) to the hotter models used by the modeling routine at the highest resolution available. The lines are very similar, suggesting that the radial velocity and $v \sin i$ but not the temperature or gravity can be extracted from the Keck data.  

\subsection{Optical Spectra}

Optical (far-red) long-slit spectra were obtained with the MMT and Gemini-North telescopes. The MMT observations were on UT Date 26 August 2012 with the Red Channel spectrograph using grating 270, using three 600 second exposures. Conditions were non-photometric. This time was allocated by NOAO as program 2012B-0233. The wavelength coverage was 6170 - 9810\AA~ with a resolution of $\sim12$\AA, but we make no use of the spectrum redder than 9250\AA~ due to strong telluric water absorption.  The Gemini-North observations (Gemini program GN-2012B-Q-105) were on UT Date 11 September 2012 with the GMOS spectrograph \citep{Hook:2004lr} using grating R831. Four 600 second exposures were taken during cloudy conditions. The wavelength coverage was 6340 to 8460\AA~with a resolution of $\sim2$\AA, but for all plots we smooth to a resolution of $\sim 10$\AA~due to low signal-to-noise. The spectra were processed using standard IRAF tasks.

\begin{figure*}
\plotone{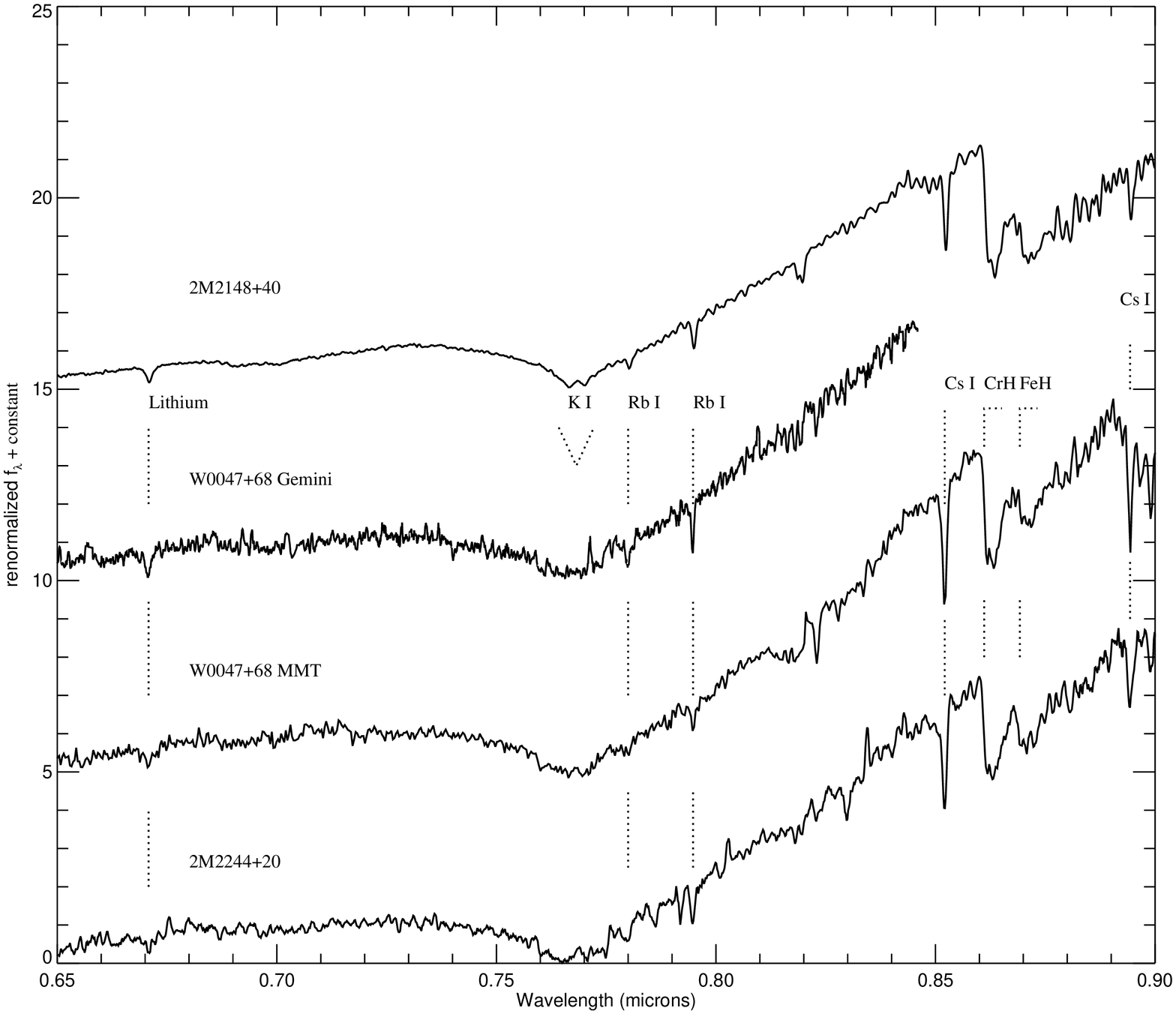}
\caption{Optical Spectra of W0047+68, along with the red L6.5 (opt) 2M2244+20 from \citet{2008ApJ...689.1295K} and the red L6 (opt) 2M2148+40 from \citet{Looper:2008lr}.  All four spectra have been smoothed.  
\label{fig-optical}}
\end{figure*}

\begin{figure*}
\plotone{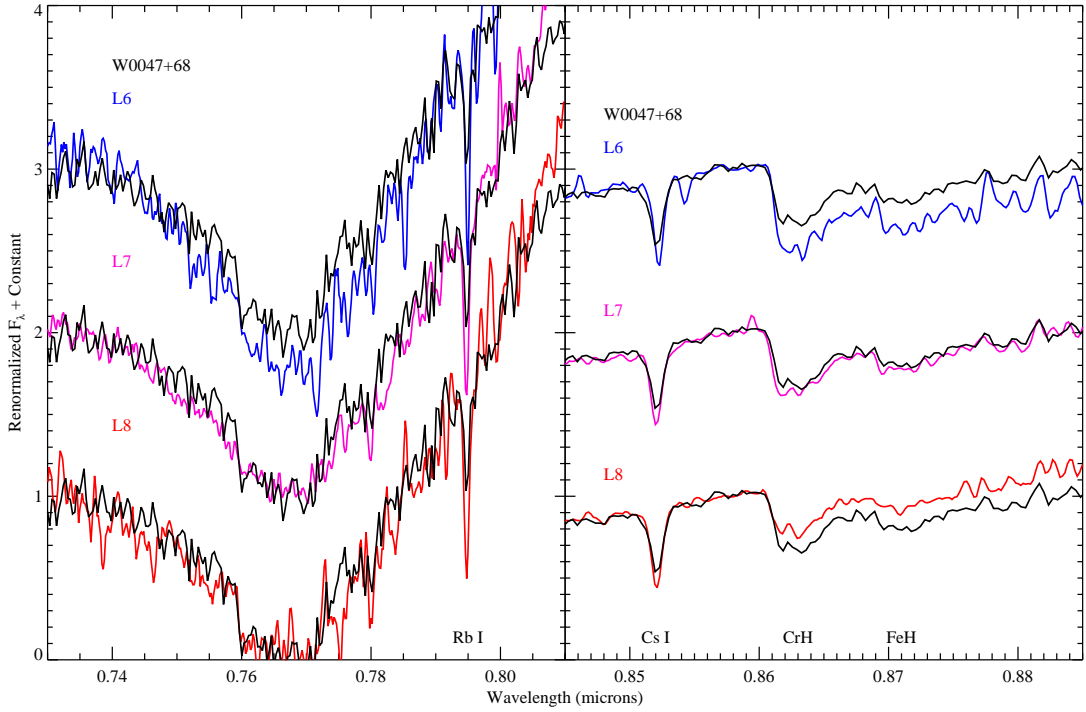}
\caption{Optical MMT spectrum of W0047+68 (black) compared to the \citet{1999ApJ...519..802K} spectra of L6, L7, and L8 standards in the \ion{K}{1}  doublet region and the CrH/FeH region; on the left, the spectra are normalized by the mean value in the range 0.72--0.81 microns and on the right, 0.855-0.860 microns. W0047+68 best resembles the L7 standard, but the \ion{Rb}{1}  and \ion{Cs}{1} lines are both weaker than expected. Note that the W0047+68 MMT spectrum shown here was taken with a different telescope and instrument, but we have smoothed the data to match resolutions.  
\label{fig-optdetails}}
\end{figure*}

\begin{figure*}
\plotone{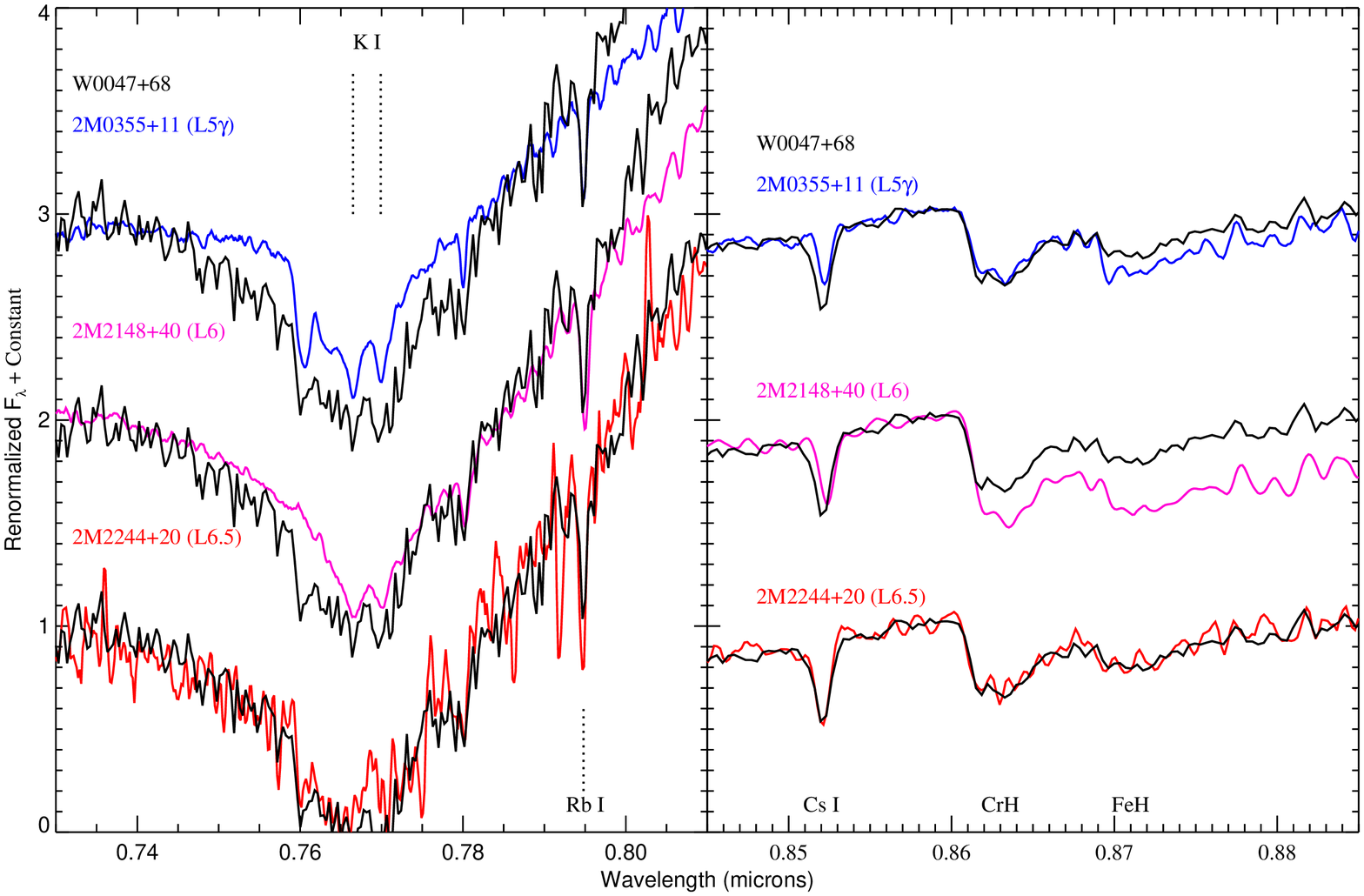}
\caption{Same as Figure~\ref{fig-optdetails}, but for the unusually red L dwarfs 2M0355+11 \citep{2009AJ....137.3345C}, 2M2148+40 \citep{Looper:2008lr} and 2M2244+20 \citep{2008ApJ...689.1295K}.  W0047+68 appears later/cooler than 2M0355+11 and 2M2148+40 but has a nearly identical spectrum to 2M2244+20.  The \ion{Cs}{1} are well-matched in 2M2244+20 and W0047+68, but the signal-to-noise is not sufficient to compare the \ion{Rb}{1} line. Again, the spectra are from different instruments so we have smoothed the data to match resolutions.
\label{fig-optdetailsred}}
\end{figure*}

The MMT and Gemini spectra are shown in Figure~\ref{fig-optical}, along with 2M2244+20, classified as L6.5 in the optical \citep{2008ApJ...689.1295K}, and 2M2148+40, classified as L7 in the optical \citep{Looper:2008lr}. The core of the \ion{K}{1} doublet and the CrH and FeH features are shown in Figure~\ref{fig-optdetails} along with the original \citet{1999ApJ...519..802K} L6, L7 and L8 standard star spectra.  The CrH and FeH features most strongly support an L7 classification, and the \ion{K}{1} doublet and overall appearance of the spectrum is consistent with this. It is evident, however, the \ion{Rb}{1} and \ion{Cs}{1} lines are much weaker in W0047+68 than in the standards or typical L6-L8 dwarfs \citep{1999ApJ...519..802K,2000AJ....120..447K}, suggesting a classification of ``L7 peculiar" is appropriate. In Figure~\ref{fig-optdetailsred}, we show the same comparison to the unusually red L dwarfs 2M0355+11 \citep{2009AJ....137.3345C}, 2M2148+40 \citep{Looper:2008lr} and 2M2244+20 \citep{2008ApJ...689.1295K}. These spectra were classified as ``L5$\gamma$", L6, and L6.5 respectively.  The weak \ion{Cs}{1} is also evident in 2M2148+40 and 2M2244+20.  W0047+68's alkali lines are stronger than in the low gravity 2M0355+11, but 2M0355+11 is a warmer type. Given our near-infrared classification of INT-G, the atomic line peculiarities hint that W0047+68 might be classified as L7$\beta$ in the optical, but we believe that extending the \citet{2009AJ....137.3345C} system to such a late-type object would be premature without more objects.  Both W0047+68 observations show lithium absorption, but no detectable H$\alpha$ emission.  

\subsection{Astrometry}

W0047+68 is being observed as part of the USNO optical parallax program, which is described in \citet{dahn}.  Based on 23 frames taken in 2 seasons, the preliminary absolute parallax is $82 \pm 3$ mas with relative proper motion $\mu = 434 \pm 3$ mas yr$^{-1}$ at position angle $\theta = 117.0 \pm 0.5^\circ$.  We derive the U,V, W space velocity of W0047+68 using the observed radial velocity and astrometry and the IDL code {\it gal\_uvw.pro}, except that we adopt the traditional astronomer convention that U is positive towards the Galactic Center. Note that W is strongly constrained ($\pm 0.5$ \kms) by the astrometry because of the position on the sky ($l=122.5$, $b=+05.2$).  We note that the absolute magnitude at H-band agrees with that of typical L7 dwarfs \citep{Dupuy:2012fk}; W0047+68 is fainter (``sub-luminous") at shorter wavelengths (J-band) and brighter (``over-luminous") at longer wavelengths (K, WISE).

Few trigonometric parallaxes of unusually red L5 or later dwarfs exist, complicating comparisons of W0047+68 to other objects. We are able to also report a preliminary USNO optical parallax of 2M2148+40, which was classified as L6 in the optical and L6.5 peculiar in the near-infrared by its discovery paper \citep{Looper:2008lr} and is L6 {\sc FLD-G} in the \citet{2013ApJ...772...79A} system.  
The observed absolute parallax is  $124.07 \pm 0.55$ mas, with proper motion of $901.9 \pm 0.3$ mas~yr$^{-1}$ at angle $59.1 \pm 0.1$ degrees (Table~2). The distance of $8.060 \pm 0.036$ parsecs places it just outside the ``8 parsec" sample.  The radial velocity is unknown, but because of the position on the sky ($l=89.0, b=-10.5$), this mainly affects the V component of space motion.  U is strongly constrained: If we adopt $v_{rad} = 0 \pm 40$ \kms, then U$= -34.0 \pm 0.5$ \kms, and W$= -5.6 \pm 7$ \kms. This rules out membership in any of the known very young nearby moving groups. 2M2148+49 is intrinsically fainter at J-band than ordinary field L6 dwarfs and brighter at WISE bands.  

\subsection{Photometry\label{sec-phm}}

W0047+68 was observed as part of normal survey operations of the Pan-STARRS wide-field optical/NIR PS1 survey \citep{2010SPIE.7733E..12K}.  We report AB magnitudes for the $y$ and $z$ bands, which are described in \citet{2012ApJ...750...99T}.  The source was not detected in the shorter wavelength bands. WISE and 2MASS photometry were previously reported in Paper I.  

We monitored W0047+68 in J-band using the FLAMINGOS camera \citep{2003SPIE.4841.1611E} on the Kitt Peak 2.1m telescope in October 2012 (NOAO-Program 12B-0233) on three consecutive nights for four hours per night, after the main target for the program had set. We find no evidence of variability at the level of 0.05 magnitudes or greater.  

Although unobservable with WISE, many of the directly imaged planets have been observed at $L'$ and $M'$.  W0047+68 has not been observed through these filters, but we may estimate the magnitudes using WISE photometry and other L dwarfs.  
\citet{2007ApJ...655.1079L} measured $M'=11.90 \pm 0.03$ for 2M2244+20. Scaling by the W2 magnitudes, this suggests $M'=11.04$ for W0047+68 and $M_{M'} = 10.6$. 

\subsection{Luminosity\label{sec-lum}}

\begin{figure}
\plotone{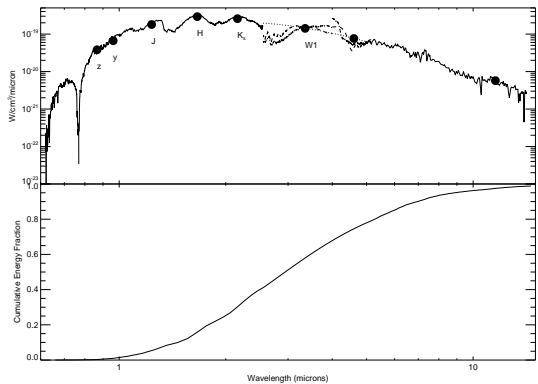}
\caption{Top: The observed spectral energy distribution. The Pan-STARRS, 2MASS and WISE photometry as shown as solid circles. The solid lines show the optical (MMT) and near-infrared (IRTF) spectra of W0047+68 and the Spitzer IRS spectrum of 2M2244 normalized to match W0047+68. The dotted lines show a very-thick-cloud (A-type) model \citep{Madhusudhan:2011yq} with $T_{eff}=1200$K, $\log g =4.0$, a BT-Settl model \citep{2013MSAIS..24..128A} with $T_{eff}=1400$K, $\log g =4.0$, and a blackbody (1400K) spectra normalized to match the W1 and W2 photometry; these are allowed to have discontinuities at 2.55, 4.00, and 5.15 microns.  Bottom: The cumulative spectral energy fraction. Half the total energy of W0047+68 is emitted shortward of 2.9 microns.  
\label{fig-sed}}
\end{figure}

We can obtain the bolometric magnitude of W0047+68 by integrating the spectral energy distribution, improving the estimate from Paper I using our new data. The near-infrared component is well-constrained by the Paper I spectrum normalized to agree with the 2MASS photometry. In the optical, we use the MMT spectrum. There is negligible contribution at less than 0.6 microns. We use the Spitzer IRS spectrum of 2M2244+20 \citep{Stephens:2009qy} renormalized by a factor of 2.23 (from the relative WISE W1 and W2 photometry) in the range 5.15 to 14.15 microns and a Rayleigh-Jeans tail beyond that. This spectral energy distribution is shown in Figure~\ref{fig-sed}. The wavelength range between 2.55 and 5.15 microns is constrained by the WISE photometry, but we require higher resolution to account for the non-smooth spectrum. We consider the A-type (1200K) and AE-type (1300K--1500K) models of 
\citet{Madhusudhan:2011yq} and the preliminary BT-Settl (1300K--1500K) of \citet{2013MSAIS..24..128A}.  In each case, we integrate the model spectra over the range 2.55 to 4.00 microns, normalized to the WISE W1 photometry, and the range 4.00 to 5.15 microns, normalized to the WISE W2 photometry. (Because the observed colors are redder than the models, we allow discontinuities.)  
Averaging the three best-fitting near-infrared models (A 1200K, AE 1300K, and BT-Settl 1400K, all $\log g=4.0$) we obtain 
an apparent bolometric magnitude of $16.30$ with an uncertainty of $\pm 0.02$ magnitudes due to the spread in models, all of which feature deep water bands. If blackbodies without absorption bands are used for this region, the resulting bolometric magnitude is up to 0.05 magnitudes brighter; while an observed AKARI spectrum of an ordinary L8 dwarf \citep{Sorahana:2012ys,Sorahana:2013lr} normalized to the W1 photometry is 0.02 magnitudes fainter; other, higher-gravity bluer models are within this range of uncertainty.  There is an additional uncertainty from the uncertainty in the 2MASS and WISE photometry and other spectra, which we estimate is $\pm 0.025$ magnitudes. The derived W0047+68 luminosity is $(3.58 \pm 0.29) \times 10^{-5} L_\odot$. The cumulative contribution to the luminosity is also shown in Figure~\ref{fig-sed}: half the energy is emitted redwards of 2.9 microns. Following the same procedure for 2M2148+40, but with the BT-Settl 1500K model, yields an apparent bolometric magnitude of $15.12 \pm 0.03$ and luminosity $L=(4.64 \pm 0.14)  \times 10^{-5} L_\odot$.

\section{Properties of W0047+68}

\subsection{AB Dor Moving Group Membership\label{sec:membership}}

The spectra of W0047+68 include a number of age indicators.  The most conservative age estimate for W0047+68 comes from the fact that we detect lithium: For the observed luminosity, this indicates its age is less than one billion years old \citep{2000ApJ...542..464C}. The near-infrared spectroscopy, however, points to an even younger age, less than that of most lithium L-type brown dwarfs. \citet{Thompson:2013fk} suggested that W0047+68 is ``low-gravity object and hence young," based on their analysis of a moderate resolution spectrum similar to the one we have obtained. Our near-infrared surface gravity classification of INT-G (Section~\ref{sec:nearir}) indicates an age of $\sim 100-150$ Myr for W0047+68, but not $<50$ Myr, though this system is primarily based upon warmer objects. Many of the youngest stars are found to be members of moving groups.

\begin{figure*}
\includegraphics*[scale=.75,angle=90]{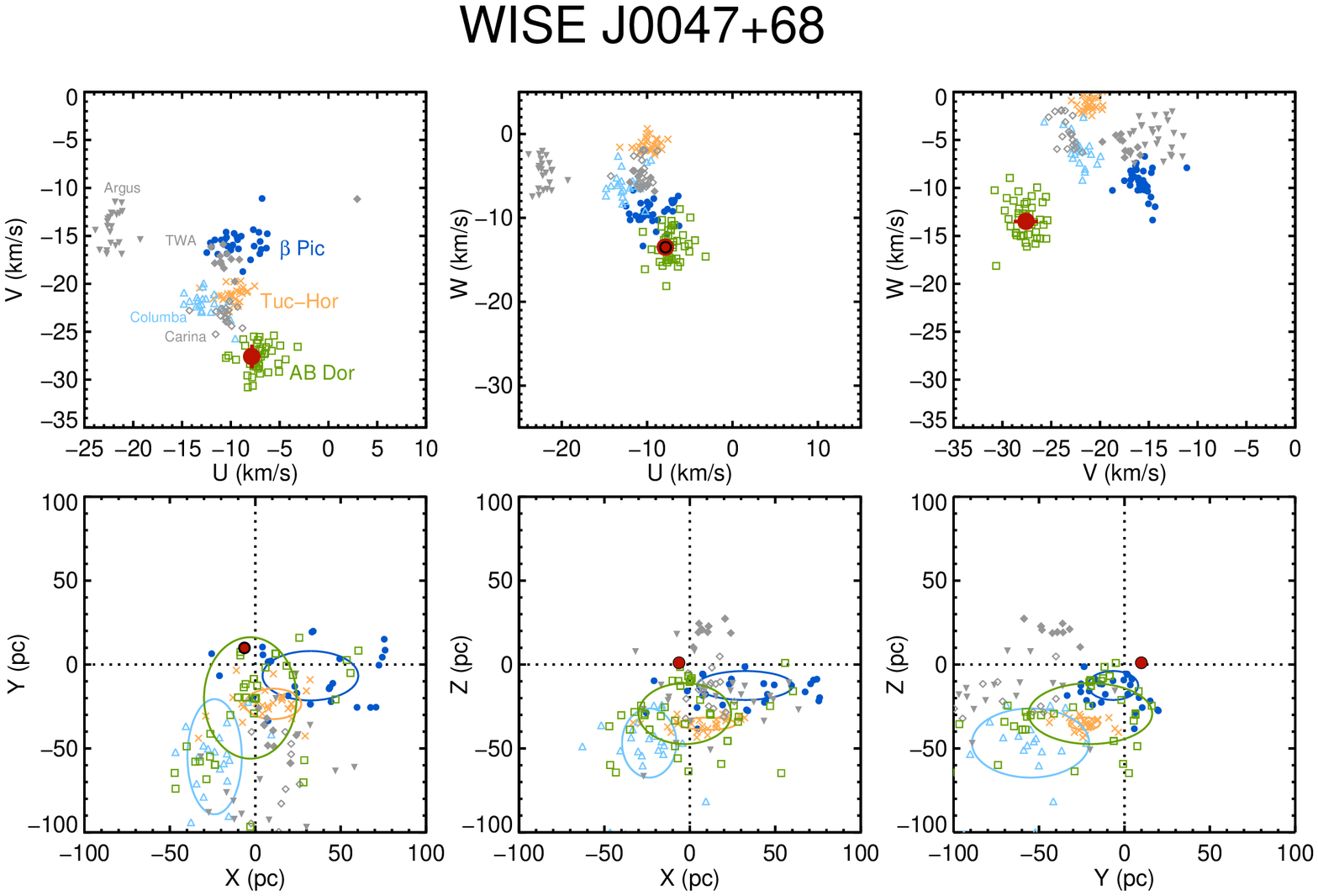}
\caption{Comparison of the position and velocities of W0047+68 (red filled circle) to nearby young associations \citep{2008hsf2.book..757T}. Although at the edge of the AB Dor Association the kinematic agreement is excellent. The size of the W0047+68 symbol is the one sigma uncertainty. \citet{Luhman:2005fj} place the Pleiades at
U$= -6.6 \pm 0.4$ \kms,V$= -27.6 \pm 0.3$ \kms, $-14.5 \pm 0.3$ \kms, close to both the center of the ABDMG distribution and W0047+68 itself in velocity space, but in physical space the cluster is located at X$=-119$ pc,Y$=28$ pc, and Z=$-53$ pc.  
\label{fig-xyzuvw}}
\end{figure*}

The three-dimensional space velocity (Figure~\ref{fig-xyzuvw}) is consistent with the young AB Dor Moving Group \citep{2004ApJ...613L..65Z,2008hsf2.book..757T}.\footnote{In Paper I, we noted that the direction of the proper motion of W0047+68 agreed with the very young ($\sim 12$ Myr) $\beta$ Pic Moving Group. The parallax rules $\beta$ Pic Moving Group membership out.} Most ABDMG members are in the southern hemisphere, but in fact the group is quite loose and includes northern hemisphere members. Given the parallax, proper motion, radial velocity and the lithium age constraint, we find that the BANYAN II model \citep{2014ApJ...783..121G} assigns a ABDMG membership probability of 99.96\% to W0047+68. As seen in Figure~\ref{fig-xyzuvw}, W0047+68 lies fairly close to the group in physical space, and the peculiar velocity necessary to move W0047+68 from the nucleus of the ABDMG to its present position is $<1$ \kms. Furthermore, as discussed in detail in Section~\ref{sec:abdmgbrown}, mid-L dwarf ABDMG candidate members, including the companion CD$-$35 2722B \citep{2011ApJ...729..139W} which is a definite member, are unusually red in $J-K$ color.
  
Therefore, on the basis of three indicators (three-dimensional space motion, spectroscopic surface gravity, and near-infrared color), we propose that W0047+68 is a {\it bona fide} member of the ABDMG. One could ask if it is plausible that an ABDMG brown dwarf member is so close to the Sun. \citet{2008hsf2.book..757T} list eight ABDMG M dwarf members within 16 pc of the Sun; and \citet{2012ApJ...753..156K} finds the ratio of brown dwarfs to stars is 1:6, so the presence of one or two nearby ABDMG brown dwarfs is not surprising.

\subsection{Age and Metallicity for W0047+68\label{sec:age}}

We can leverage current knowledge of the ABDMG, which is based on a wide variety of stellar masses and spectral types, to understand the properties of W0047+68 and the origin of its peculiar spectrum.  
Although early discussions of ABDMG \citep{2004ApJ...613L..65Z} assumed an age of $\sim 50$ million years, current evidence points to an age of $\sim 100-125$ Myr \citep{Luhman:2005fj,Barenfeld:2013qy}. Indeed, \citet{Luhman:2005fj}  argue that ABDMG is ``roughly co-eval" with the Pleiades and is physically related given the similar space velocities, an argument supported by \citet{2007MNRAS.377..441O}. This age is in excellent agreement with our spectroscopic estimate of surface gravity intermediate between the youngest association and older field brown dwarfs. Of course, even the age of the rich, well-studied Pleiades is still somewhat uncertain (see \citealt{Soderblom:2009kx}), but W0047+68's age constraint is much better than that of most known brown dwarfs.

The relative importance of surface gravity and composition is debated in both unusually red brown dwarfs and massive gas giant planets. Crucially, the composition of the ABDMG is also known to be near solar.  \citet{2007MNRAS.377..441O} compiled recent literature measurement of 15 ABDMG members and found a mean $[Fe/H]= - 0.02 \pm 0.02$.   \citet{2009A&A...501..965V} measured mean $[Fe/H] = - 0.01]$ for 12 ABDMG members, with a standard deviation of 0.09, but after applying a correction for a trend in effective temperature found a mean $[Fe/H]= 0.04$ with standard deviation of 0.05. They also reported that for their entire young star sample the mean values $[Si/H ]=-0.11$ and $[Ni/H] = 0.01$. \citet{2012MNRAS.427.2905B} reported a mean $[Fe/H] = 0.10 \pm 0.03$ for the five stars in the ABDMG, and found that ten elements (Na, Mg, Al, Si, Ca, Ti, Cr, Fe, Ni and Zn) have abundance patterns that are solar and ``typical" of the local Galactic thin disk. \citet{Barenfeld:2013qy} claim a somewhat lower $[Fe/H] = 0.02 \pm 0.02$ for the ABDMG nucleus and measure seven other elements (Na, Mg, Al, Si, Cr, Mn, Ni) to be close to solar values. Since these $[Fe/H]$ values are characteristic of the bulk of the local thin disk, the vast majority of field L dwarfs must have similar values. Whether the scatter in values is a matter of observational uncertainties or true scatter could be debated. \citet{Barenfeld:2013qy} argue that they see real scatter in composition in the ABDMG ``stream" members that therefore many of the more distant ABDMG objects are not from a common origin. Nevertheless, even these objects are all near solar in $[Fe/H]$ and other elements, as is the Pleiades \citep{Soderblom:2009kx}, so we conclude that W0047+68's composition is constrained to be near-solar even if the ABDMG stream is not truly homogeneous. This constraint includes the key elements involved in cloud formation: Fe and Al (along with O) are the major elements for high-temperature condensates (i.e., in the deeper atmospheric layers of W0047+68) and Mg and Si (and O) at moderate temperatures characteristic of the upper atmospheric layers \citep{Marley:2002fk,Helling:2008mz,2010ApJ...716.1060V} . 

\begin{figure*}
\includegraphics[scale=.75,angle=90]{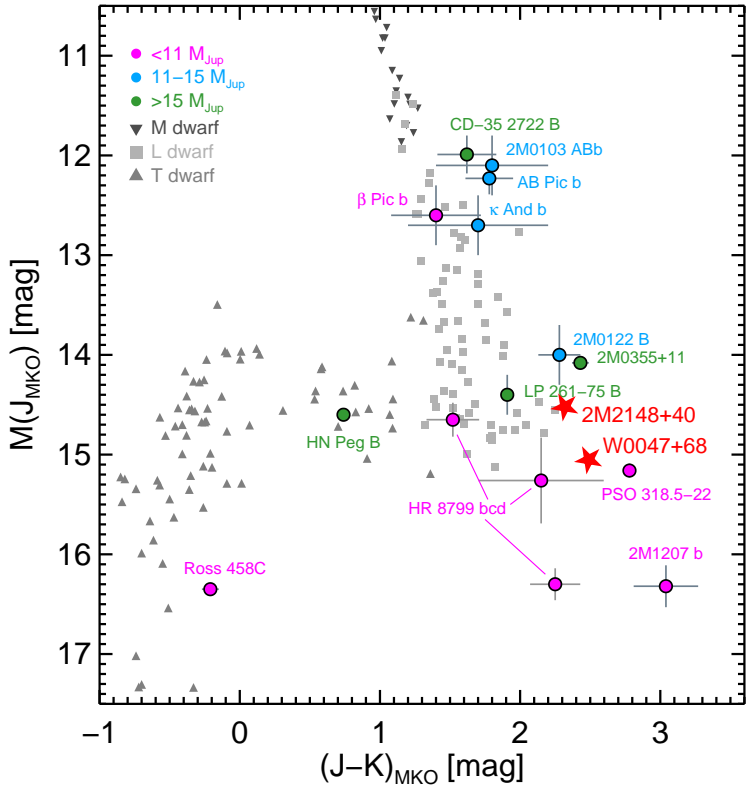}
\caption{Color-magnitude diagram comparing W0047+68 and 2M2148+40 to ordinary field L and T dwarfs, very-low-mass brown dwarfs, and exoplanets compiled by \citet{2013ApJ...777L..20L}. 
\label{fig-hr}}
\end{figure*}

Using the \citet{2000ApJ...542..464C} models and the observed luminosity, W0047+68 would have $\log g \approx 4.5$ and $M \approx 0.018 M_\odot$ for age 120 Myr. Similar values are found with the \citet{1997ApJ...491..856B} models. The low surface gravity and solar metallicity suggests that the thick clouds of W0047+68 are a normal feature of a young brown dwarfs.  In Figure~\ref{fig-hr}, we show the color magnitude diagram for brown dwarfs and planetary mass objects \citep{2013ApJ...777L..20L}, updated with W0047+68  and 2M2148+40.  W0047+68 lies close to PSO J318-22, adding to the evidence of a low-gravity sequence that extends redward of the normal L/T brown dwarf sequence.

\subsection{Effective Temperature\label{sec-temp}}

In Paper I, we found that although models from different  groups agreed that unusually thick clouds were needed to fit the
near-infrared spectrum, the predicted effective temperatures ranged from as high as 1600K to as low as 1100K.  With our measurement of the luminosity, we can make another estimate of the effective temperature using theoretical predictions of the radius.  We find:

\begin{equation}
T_{eff} = (1413 {\rm K} \pm 28 {\rm K) } \left({0.1 R_\odot}\over{R}\right)^{0.5}   
\end{equation}

The highest temperature and smallest radius for W0047+68 can be obtained by using the oldest evolutionary models in which lithium has not been depleted. This would allow $T_{eff} \approx 1450$K and an age of $\sim1$ billion years, with $\log g \approx 5$. ABDMG membership implies a larger radius and lower temperature. Adopting an age of 125 Myr for ABDMG  membership, \citet{1997ApJ...491..856B} have $M_\odot = 0.020 M_\odot$ and $R = 0.117 R_\odot$; the \citet{2000ApJ...542..464C} are $M = 0.018 M_\odot$ and  $R = 0.124 R_\odot$. Correspondingly, our best estimate of W0047+68's effective temperature is $T_{eff} \approx 1270$--$1300$K with $\log g \approx 4.5$.   \citet{Burrows:2011lr} report that high metallicity or thick cloud models each lead to an increase in the radius over ``standard" models, which would lower the temperature below 1400K even if W0047+68 is old.  

Given an effective temperature of $\sim 1300$K, W0047+68 shows that intermediate surface gravity brown dwarfs have a much different
L/T type transition. For older, high gravity field brown dwarfs, this effective temperature is normally associated with the early T dwarfs \citep{Cushing:2008kx,Stephens:2009qy} and certainly with spectral types that already begun the move to the blue in $J-K$.  The early T dwarfs have methane bands in H-band and K-band \citep{2006ApJ...637.1067B}. No such features are seen in W0047+68, and we also did not detect the high amplitude J-band variations seen in $\sim 1300$K early-T dwarfs and attributed to patchy clouds \citep{Artigau:2009ly,Radigan:2012vn,2013ApJ...768..121A}. (However, W0047+68's low $v \sin i$ suggests we may be viewing it near pole-on.) 
We emphasize that this result depends upon the evolutionary models: W0047+68 has a luminosity typical of that of ordinary L7 dwarfs \citep{2004AJ....127.2948V,2004AJ....127.3516G}, but its young age (and theoretically larger radius) requires a lower temperature. As shown by \citet{2013ApJ...777L..20L}, the younger object PSO J318-22 is an even more extreme case, with nearly the same luminosity and spectral type as W0047+68 but a deduced temperature of $1160^{+30}_{-40}$K.  

\subsection{Comparison to Thick Cloud Models\label{sec-models}}

The properties of condensate cloud layers depend on complex processes that are difficult to model completely, including nucleation, growth, and settling \citep{Ackerman:2001fj,Helling:2008mz}. Each theoretical effort to predict synthetic spectra makes different choices in parameterizing the cloud properties. These were compared by \citet{Helling:2008uq} in test cases aimed at modeling typical field brown dwarfs. In Paper I, we discussed how theoretical models from three different groups (UCM, \citealt{2005ApJ...621.1033T}; $f_{sed}$, \citealt{Ackerman:2001fj,Marley:2002fk,Stephens:2009qy}; very thick (A-type) and intermediate thickness (AE-type) cloud models, \citealt{Madhusudhan:2011yq}) were able to fit the low-resolution W0047+68 near-infrared spectrum. In each case, models can fit the $J-K$ color and qualitatively reproduce the spectrum by adjusting a free parameter to increase the thickness of the condensate clouds. Another set of models, {\rm DRIFT-PHOENIX} \citep{2009A&A...506.1367W} was used by \citet{Witte:2011fk} to fit low-resolution near-infrared spectra of 2M2148+40 and 2M2244+20. They derived warm temperatures ($T_{eff}=1500$K) and very low gravities ($\log g =3.0$ and $3.5$~respectively), but discussed the model grid limitations for such red objects and concluded that it is was likely their temperatures are overestimated and the gravities underestimated.

\begin{figure*}
\plotone{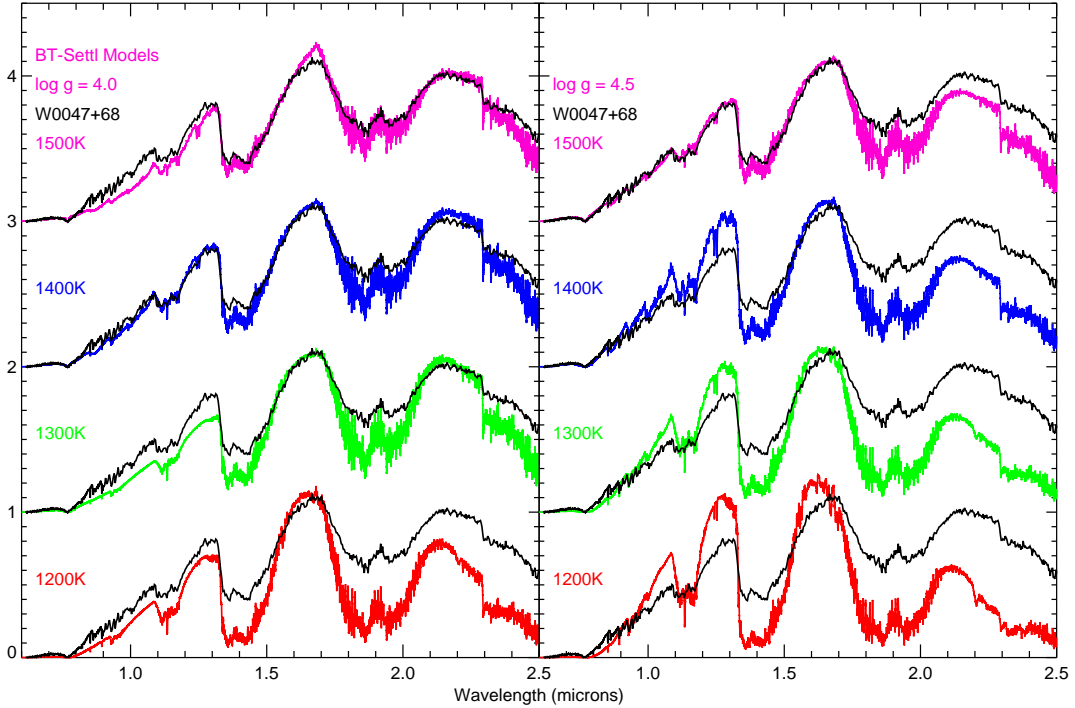}
\caption{Comparison of observed W0047+68 spectrum to low gravity ($\log g = 4.0, 4.5$) BT-Settl models: Each is normalized to agree in the 1.50-1.75 micron region. While we conclude that $T_{eff} = 1300-1430$K and $\log g = 4.5$, the models with $\log g =4.0$ do a better job of reproducing the red near-infrared spectrum.    
\label{fig-allard}}
\end{figure*}

We can now also compare to a fifth set of models. \citet{Allard:2012ul,2013MSAIS..24..128A} have presented a new model family called BT-Settl. The mixing and cloud properties are informed by two-dimensional radiation hydrodynamic (hereafter RHD) simulations \citep{2010A&A...513A..19F}. The result is that for cool L dwarfs, low-gravity objects are predicted to be redder, with thicker clouds. We show the W0047+68 observed spectrum with the model predictions in Figure~\ref{fig-allard}. While our analysis of W0047+68 favors $T_{eff} \approx 1300$K and $\log g \approx 4.5$, the BT-Settle grid requires an even lower gravity but higher temperature ($T_{eff} = 1400-1500$K, $\log g = 4.0$) to match the spectrum.  (Note that the $\log g = 3.5$ BT-Settl models are even redder and can also fit.)

It has long been recognized that $J-K$ colors are relatively poor temperature indicators for brown dwarfs, even amongst the typical field population, because of the effects of clouds \citep{Marley:2002fk,2004AJ....127.3553K,Stephens:2009qy}.  Our results again underline this issue. The Unified Cloudy Model and BT-Settl give qualitatively good fits to W0047+68, but are warmer by 100K-200K than our estimate of 1300K, while on the other hand, the very thick (A-type) and $f_{sed} =1$ models give temperatures 100-200K too low when matching the spectra. The plots in Figure~\ref{fig-allard} have arbitrary normalizations. An interesting check is to derive the radius from the near-infrared region alone instead of using our procedure in Section~\ref{sec-temp}: given a model, the observed parallax, and the observed flux-calibrated spectrum, the only adjustable parameter is the radius. The BT-Settl results when fitting over the range 1.2 to 2.5 microns with the $\log g = 4.0$ models are $0.13 R_\odot$ (1200K), $0.12 R_\odot$ (1300K), $0.94 R_\odot$ (1400K), and $0.82 R_\odot$  (1500K), which is consistent with the discussion in Section~\ref{sec-temp}. The 1500K model has a qualitatively good fit but requires a radius considerably smaller than we expect for a brown dwarf with lithium. We believe the balance of evidence is against this possibility, although it should be noted that \citet{Sorahana:2013lr} have used their own model fitting to AKARI data to argue that (ordinary) field late-L dwarfs are warmer and much smaller than predicted. \citet{Cushing:2008kx} showed that fits to limited wavelength regions can almost always work but can give much different effective temperatures than fits to the entire spectral energy distribution.  

One important conclusion from our work is that model colors should not be used to derive surface gravities because they cannot yet predict cloud properties well enough. In the cases with a full grid of surface gravities \citep{Madhusudhan:2011yq,Allard:2012ul}, one must go to $\log g =4.0$ or even less to get thick enough clouds to match in the infrared. This would in turn require that W0047+68 be below the deuterium-burning limit and very young, much younger than the best estimates of the ABDMG age. 
For example, in the \citet{Baraffe:2002mz} models at 20 Myr, a $0.007 M_\odot$ would have $\log g =4.0$ and match the observed luminosity with $T_{eff}=1200K$, while a $0.009 M_\odot$ would have $\log g = 4.1$  and $T_{eff} = 1400$K but would be 90\% more luminous than observed.  
\citet{2014A&A...564A..55M} have argued that the high altitude dust in BT-Settl models should be increased to better fit the warmer (M9.5-L3) young  brown dwarfs; the same is evidently true for W0047+68.  

The optical spectrum of W0047+68 also poses a challenge for our understanding. The spectrum is dominated by the broad wings of the \ion{K}{1} which in turn are the dominant feature that can be used for spectral typing; these are evidently saturated around 1300K even though their weakness is indicator of low gravity in warmer objects \citep{2009AJ....137.3345C}.
This was recognized by \citet{2008ApJ...689.1295K} who argued that 2M2244+20 is young yet found a normal L dwarf spectrum (shown in our Figures~\ref{fig-optical} and \ref{fig-optdetailsred}).  They speculated that ``the lower gravity [optical] spectrum will have weaker alkalis and stronger hydrides than a high-gravity L dwarf of comparable temperature."  This in turn creates ``a degeneracy at mid- and late-L types whereby low-gravity objects are very difficult or impossible to distinguish in the optical." If our interpretation of W0047+68 is correct, this must be the case because (except for the strengths of the Cs and Rb lines) it is very close to the L7 standard.  Another important factor in creating degeneracy may be additional opacity due to dust, which we know must be different in the near-infrared and may be different in the optical. \citet{Pavlenko:2000yq} discussed the optical spectra of L dwarfs and showed that the broad \ion{K}{1} line and the sharper \ion{Cs}{1} and \ion{Rb}{1} lines in the $T_{eff} = 1200-1500$K range are sensitive to surface gravity and dust extinction;  they could change the best fit effective temperature and surface gravity by increasing or decreasing them together, and additional opacity helped the quality of fits.  \citet{Lodders:1999qy}, \citet{2002Icar..155..393L} and \cite{Marley:2002fk} discuss the sensitivity of the \ion{K}{1} chemistry to sedimentation and cloud formation. Thus it may be possible to find models where the ``L7" optical spectra are relatively insensitive to gravity.  

\begin{figure}
\epsscale{0.6}
\plotone{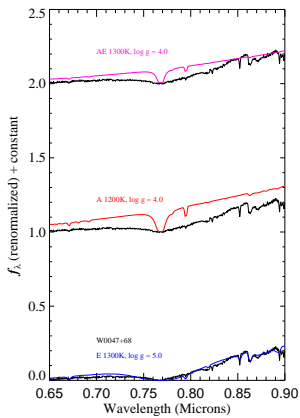}
\caption{Comparison of observed W0047+68 spectrum to \citet{Madhusudhan:2011yq} intermediate thickness clouds (AE type) and very thick cloud (A type) models with low surface gravity ($\log g =4.0$) models. These two models are normalized to the W0047 H-band region (as in Paper I). The strength of the broad potassium line is under-predicted, and the pseudo-continuum is too high, due to the overly low surface gravity. We also show a \citet{2006ApJ...640.1063B} model with 1300K $\log g =5.0$ and normal, thin ("E-type") clouds, normalized in the region 0.7-0.9 microns, which demonstrates that this model family does reproduce the main features of L dwarfs in the optical.
\label{fig-burrowsopt}}
\end{figure}

\begin{figure}
\plotone{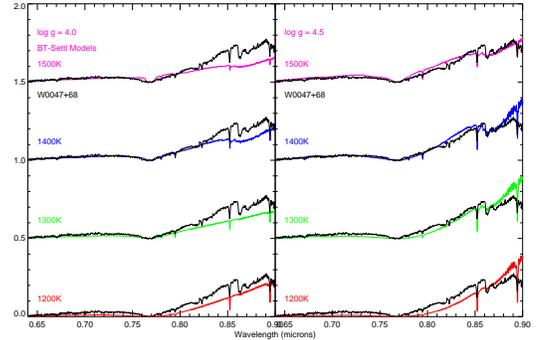}
\caption{Comparison of observed W0047+68 spectrum to low gravity ($\log g = 4.0, 4.5$) BT-Settl models: Each is normalized to agree in the 1.5-1.75 micron region. See the discussion in Section~\ref{sec-models}.
\label{fig-allardopt}}
\end{figure}

Synthetic optical spectra are available for the BT-Settl and \citet{Madhusudhan:2011yq} model families, and have not been tuned to match observations.  In Figure~\ref{fig-burrowsopt}, we compare the predicted and observed far-red spectra using the \citealt{Madhusudhan:2011yq} very-thick (A) and intermediate thickness (AE) models {\it normalized to agree in the H-band} (as in Figure~6 of Paper I). In both cases, the absorption by the broad \ion{K}{1} wings is underestimated, and as a result, the continuum is too high. Since the optical and near-infrared spectra of typical L dwarfs are fit well by the ``E type" thin cloud models and $\log g =5.0$ \citep{2006ApJ...640.1063B}, one would expect them to also fit the optical spectrum of W0047+68. Such a fit is also shown in Figure~\ref{fig-burrowsopt}. The optical BT-Settl models are shown in Figure~\ref{fig-allardopt}. There are also disagreements with the observations, with the pseudo-continuum lying too low, but the $T_{eff} =1300$K, $\log g =4.0$ model is reasonably similar in overall morphology and slope, though it lacks the hydrides; CrH and FeH are better matched with the $\log g =4.5$ models. The change in shape of the core of the \ion{K}{1} resonant doublet in the $\log g =4.0$ models from 1500K to 1300K is similar to the differences between 2M0355+11 and W0047+68 (Figure~\ref{fig-optdetailsred}), and suggests the basic picture that the core of the \ion{K}{1} feature is saturated for late-L dwarfs even for low gravities may be correct.  In both the BT-Settl and A/AE models, the $\log g=4.0$ models have no CrH or FeH features: This prediction could be tested in the $\beta$ Pic Moving Group planetary-mass-object PSO J318-22.

\subsection{AB Dor Moving Group Brown Dwarfs\label{sec:abdmgbrown}}

\begin{figure}
\plotone{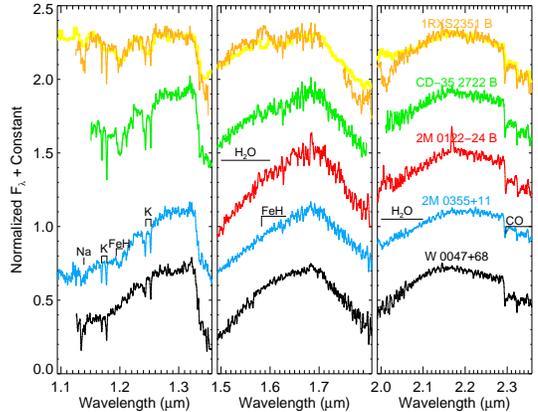}
\caption{Comparison of ABDMG L-type brown dwarfs.  Note the sharply-peaked 2M0355+11 H-band spectrum compared to the other cool L dwarf members.
\label{fig-abdmg-spectra}}
\end{figure}

There are now a number of L-type\footnote{\citet{Delorme:2012fk} have also proposed CFBDSIR 2149-0403, an unusually red T dwarf, as a candidate planetary mass ABDMG member.} brown dwarfs proposed as ABDMG members. The strongest members, CD$-$35 2722B \citep{2011ApJ...729..139W}, 2M0355+11 \citep{Faherty:2013qy,Liu:2013lr}, and W0047+68, show a surprising spread in spectral properties. 1RXS J2351+3127 B \citep {2012ApJ...753..142B} and 2MASS 0122Ð2439 B \citep{2013ApJ...774...55B} still need kinematic confirmation. The similarity of 2M2244+20 and W0047+68 supports ABDMG membership for 2M2244+20; using the W0047+68 absolute magnitudes for 2M2244+20 implies a distance consistent with ABDMG tangential motion \citep{2014ApJ...783..121G}. The signal-to-noise on the 2M2244+20 spectrum is too low for a reliable measurement of lithium.   
 In Figure~\ref{fig-abdmg-spectra} we plot the intermediate resolution near-infrared spectra of 1RXS J2351+3127 B, CD$-$35 2722B, 2MASS 0122Ð2439 B, 2M0355+11 \citep{Liu:2013lr}, and W0047+68.

\citet{2013ApJ...772...79A} have discussed the discrepancies between 2M0355+11 (L3 VL-G), and CD$-$35 2722B, which is L3 INT-G and a companion to one of the nucleus AB Dor members. Most notably, 2M0355+11 is more strongly peaked in the H-band. 
Despite the very similar $J-K_s$ colors of W0047+68 and 2M0355+11, the surface gravity sensitive features and very different, and W0047+68 appears to be cooler. 2M0355+11 is also a magnitude brighter in the color-magnitude diagram (Figure~\ref{fig-hr}).  The Pleiades is a similar age and space motion, and infrared spectroscopy shows early L dwarfs in the Pleiades are unusually red \citep{Bihain:2010fk}.  \citet{2013arXiv1307.7153A} report that these spectra can be classified VL-G or INT-G but show a spread in properties. Unfortunately, no late-type L dwarfs in the Pleiades are yet known, but our adopted age of ABDMG implies that W0047+68-like members should be found in deep Pleiades proper motion searches. Together, the ABDMG moving group brown dwarfs form an empirical intermediate surface gravity ($\log g = 4.5$) sequence.   

One puzzle is that 2M0355+11 has two very obvious indicators of low surface gravity  -- the very peculiar optical spectrum (L5$\gamma$), a very sharply peaked H-band -- while W0047+68 has more subtle features. The inconsistency of the H-band shape as a gravity indicator is discussed in detail by \citet{2013ApJ...772...79A}.  In regards to this pair of objects, three possibilities merit discussion.
One possibility is that both 2M0355+11 and W0047+68 are normal ABDMG members whose differences are simply a consequence of temperature. \citet{2013ApJ...777L..20L} have found that the luminosity of 2M0355+11 is $\log L/L_\odot = -4.23 \pm 0.11$, i.e., $\sim 60\%$ more luminous than W0047+68. This suggests it is $\sim 170$K warmer than W0047+68, consistent with the two spectral types earlier classification in the optical. The optical observed differences in the \ion{K}{1} core (Figure~\ref{fig-optdetailsred}) appear qualitatively similarly to the BT-Settl offsets between the 1500K/1400K models and the 1300K models for $\log g = 4.0$;  the BT-Settl low-gravity models are most ``peaky" at 1500K but H-band becomes more rounded for lower temperatures.  The A/AE models are also not ``peaky" at H-band for 1300K.  A less likely possibility is both 2M0355+11 and W0047+68 are ABDMG members, but 2M0355+11 is actually a lower mass object very close to the deuterium-burning limit, resulting in a higher luminosity and temperature than W0047+68 but a lower surface gravity. This kind of complex behavior near the deuterium-burning limit at ages up to $\sim 120$ Myr appear in all published model tracks and have recently been discussed again theoretically by \citet{2011ApJ...727...57S} and observationally by \citet{Allers:2010lr} and \citet{2013ApJ...774...55B}.  Whether or not this suggestion matches in detail depends on the age of ABDMG and the exact behavior of model tracks near the deuterium-burning limit.  
The third possibility is that either, or both, are not ABDMG members. Of course, if 2M0355+11 has a lower gravity than it is a {\it younger} interloper, perhaps $\sim 30-50$ Myr old.  This may not be as unlikely as it seems.  Pre-Hipparcos work identified a ``Local Association" (also known as the ``Pleiades Super-Cluster" or ``Pleiades Moving Group") including many young stars in the age range $\sim 50 - 150$ Myr \citep{Eggen:1992qy,Jeffries:1995uq}, with a dispersion of $\pm 10$ \kms.   The ABDMG is part of this broader association.   
 \citet{2001MNRAS.328...45M} identified many young stars with Local Association motions, and found kinematic ``substructure" as well as a fairly large spread in velocities. \citet{Makarov:2003uq} confirmed the Local Association streaming motion in a X-ray selected sample. 
\citet{2008A&A...490..135A} (see their Figure 13) find many young ($<100$ Myr) stars concentrated in this broad region of U-V space.
More controversially, \citet{2006ApJ...643.1160L} argued that the AB Dor group could be divided into a $\sim 30-50$ Myr-old subgroup and a $\sim 80-120$ subgroup. (There are also, of course, old stars with similar motion so kinematics alone cannot prove youth. In a non-age-selected Hipparcos sample, \citet{Bovy:2009qy} find that $\sim 10$\% of stars can be associated with a Pleiades moving group.)

\section{2M2148+40: An older dusty L dwarf}

In Paper I, we remarked that in our low-resolution near-infrared spectrum W0047+68 resembled a slightly cooler (or one subclass later) version of 2M2148+40. \citet{Looper:2008lr} discovered this brown dwarf and argued that spectroscopic and kinematic evidence suggests it is not young, but is rather an older brown dwarf whose spectral peculiarities might be explained by high metallicity. (However, they did find 2M2148+40 has lithium with equivalent width $12.1 \pm 0.6$\AA, see Figure~\ref{fig-optical}.) 
At moderate resolution significant differences from W0047+68 are revealed. \citet{2013ApJ...772...79A} found that its moderate-resolution infrared spectrum lacked signs of low surface gravity or youth, and classified it as L6 {\rm FLD-G}. This spectrum is compared to W0047+68 in Figure~\ref{fig-irallers}. With a reliable, precise distance, we can now revisit the kinematic arguments. Our measurement of the U-component of its space motion (-34 \kms) is not consistent with any known young moving groups, but is not unusual amongst a sample of stars that are $\sim 0.5-2$ billion years old (such as dMe dwarfs or $M_V <4$ main sequence stars; \citealt{2002AJ....124.2721R}, their Figure 15). 
\citet{2004A&A...420..183A} measured $[Fe/H]$ of all (118) stars more luminous than $M_V = 6.5$ mag within 14.5 pc from the Sun and found and 2-5\% have $[Fe/H] > +0.25$. The most metal-rich star in the same is \object{HD 125072} which has (U,V,W) $=$ ($-37.9, -17.1, -34.8$) \kms, proving 2M2148+40's U motion cannot be used to rule out high metallicity.  Indeed, the U velocity is consistent with the most common velocities of the very-metal-rich planet-host stars in the solar neighborhood, which are thought to be scattered from more inner regions of the Galactic Disk  \citep{2007A&A...461..171E}. \citet{2008A&A...480..889S} find that the nearby star forming regions are near-solar metallicity in both $[Fe/H]$ and $[Si/Fe]$, and argue that high metallicity stars originate in the inner galaxy, and \citet{1996A&A...314..438W} show that orbital diffusion of metal-rich stars explains the presence of high metallicity stars locally.  \citet{1977A&A....60..263W} find the typical change in the Galactic orbit radius will be 0.7--0.9 kpc in 0.5-1.0 Gyrs, the oldest ages consistent with the presence of lithium, but presumably 2M2148+40 would be an example of an unusually large change.  

Given the observed luminosity, $T_{eff} = 1507 \pm 11$K for $R= 0.1 R_\odot$. Applying the DUSTY \citep{2000ApJ...542..464C} evolutionary models to this luminosity and the observed presence of lithium, 2M2148+40 would have $M \approx 0.04  - 0.055 M_\odot$ and $\log g = 5.0 - 5.2$ for ages 0.5 to 1.0 Gyr. Our luminosity for 2M2148+40 is similar to those of ``normal" field L6 and L7 dwarfs \citep{2004AJ....127.2948V,2004AJ....127.3516G}, so we necessarily derive similar effective temperatures. In summary, the new astrometry is consistent with the existing paradigm that 2M2148+40 has a normal surface gravity and effective temperature for a lithium L6-type dwarf, so that high metallicity (or else some other unknown factor) is needed to explain its unusually thick condensate clouds. 

\section{Conclusions}

Our new measurements of W0047+68 confirm the view that it is an intermediate gravity ($\log g \approx 4.5$), young brown dwarf in the ABDMG.  Relatively bright with a reliable age, metallicity and luminosity, it is a key benchmark for characterizing the effects of clouds in substellar atmospheres. No theoretical model is fully satisfactory in matching both the optical and near-infrared spectra and luminosity, but many features are explained and the need for increased mineral clouds is clear. The low effective temperature ($\sim 1300$K) of W0047+68 reinforces the finding that the standard L/T transition sequence for field brown dwarfs ($\log g \approx 5$) does not apply at intermediate gravities ($\log g=4.5$), just as is already known for the low gravities of planetary mass objects ($\log g \approx 4.0$), and the near-solar metallicity of the ABDMG demonstrates that gravity alone is responsible for the differing cloud properties. The luminosity and temperature of the older L dwarf 2M2148+40 is warmer than the L/T transition, but another factor, presumably metallicity, places it in the redder sequence of the color-magnitude diagram among the young objects. 

\acknowledgments

We are grateful to the USNO CCD parallax team for allowing us to use the preliminary parallax measurements. 
We thank Eric Mamajek for discussions of local associations; Kimberly Aller for examining Pan-STARRS data; Denise Stephens, Satoko Sorohana, and Dagny Looper for providing electronic copies of their data; and France Allard and Adam Burrows for sharing their grids of synthetic spectra. This research has benefitted from the SpeX Prism Spectral Libraries, maintained by Adam Burgasser at \url{http://www.browndwarfs.org/spexprism}

Some of the observations reported here were obtained at the MMT Observatory, a joint facility of the Smithsonian Institution and the University of Arizona. MMT telescope time was granted by NOAO, through the Telescope System Instrumentation Program (TSIP). TSIP is funded by NSF. Based on observations obtained at the Gemini Observatory, which is operated by the Association of Universities for Research in Astronomy (AURA) under a cooperative agreement with the NSF on behalf of the Gemini partnership: the National Science Foundation (United States), the Science and Technology Facilities Council (United Kingdom), the National Research Council (Canada), CONICYT (Chile), the Australian Research Council (Australia), CNPq (Brazil) and CONICET (Argentina). Some of the data presented herein were obtained at the W.M. Keck Observatory, which is operated as a scientific partnership among the California Institute of Technology, the University of California and the National Aeronautics and Space Administration. The Observatory was made possible by the generous financial support of the W.M. Keck Foundation. The Pan-STARRS1 surveys have been made possible by the Institute for Astronomy, the
University of Hawaii, the Pan-STARRS Project Office, the institutions of the Pan-STARRS1 Science
Consortium (http://www.ps1sc.org), NSF, and NASA. This research has made use of NASA's Astrophysics Data System, the VizieR catalogue access tool, CDS, Strasbourg, France, and the NASA/ IPAC Infrared Science Archive, which is operated by the Jet Propulsion Laboratory, California Institute of Technology, under contract with the National Aeronautics and Space Administration. IRAF is distributed by the National Optical Astronomy Observatory, which is operated by the Association of Universities for Research in Astronomy (AURA) under cooperative agreement with the National Science Foundation.

\bibliography{./w0047red}


\begin{deluxetable}{lcl}
\tablewidth{0pc}
\tabletypesize{\footnotesize}
\tablenum{1} \label{tab1}
\tablecaption{WISEP J004701.06+680352.1}
\tablehead{
\colhead{Parameter} & 
\colhead{W0047+6803} & 
\colhead{Remarks}}
\startdata
Sp. Type & L7 INT-G & Infrared,  \citet{2013ApJ...772...79A} system\\
Sp. Type & L7 pec & Optical,  \citet{1999ApJ...519..802K} system\\
$\pi_{abs}$ & $82 \pm 3$ & mas \\
$\mu_{rel}$ & $434 \pm 3$&  mas/yr \\
$\theta$ & $117.0 \pm 0.5$ & degrees, East of North \\ 
$v_{rad}$ & $-20.0 \pm 1.4$ & \kms \\
$v \sin i $ & $4.3 \pm 2.2$ & \kms \\
$z$ & $18.95 \pm 0.03$ & AB mag, Pan-STARRS \\
$y $ & $18.12 \pm 0.03$ & AB mag, Pan-STARRS\\
U & $-8.6 \pm 1.0$ & \kms \\
V & $-27.8 \pm 1.3$ &\kms\\
W & $-13.6 \pm 0.5$ & \kms\\ 
$M_J$ & $15.17 \pm 0.11$ & 2MASS \\
$M_H$ & $13.54 \pm 0.09$ & 2MASS \\
$M_{Ks}$ &$12.62 \pm 0.09$ & 2MASS \\
$M_{W1}$ &$11.47 \pm 0.08$ & WISE \\
$M_{W2}$& $10.82 \pm 0.08$ & WISE \\
$M_{W3}$& $9.71 \pm 0.10$ & WISE \\
$m_{bol} $ & $16.30 \pm 0.03$ & Section~\ref{sec-lum}\\
$L/L_\odot$ & $(3.58 \pm 0.29) \times 10^{-5}$ & Section~\ref{sec-lum}\\
$BC_J$ & $0.70 \pm 0.08$ & \\
$BC_K$ & $3.25 \pm 0.04$ & \\
$M_J$ (MKO) & $15.05 \pm 0.11$ & Color term from spectrum, See Paper I \\
$M_H$ (MKO) & $13.65 \pm 0.09$ & See Paper I \\
$M_K$ (MKO) & $12.56 \pm 0.09$ & See Paper I \\
$M_{M'}$ & 10.6 & Using 2M2244, see Section~\ref{sec-phm}\\ 
\cutinhead{If AB Dor Association member} 
$T_{eff} $ & $\sim 1270$ & K, see Section~\ref{sec-temp}\\
Age & $\sim 120$ & Myr\\
Mass & $\sim 0.018$ & $M_\odot$ \\
$\log g $ & $\sim 4.5$ & \\
\cutinhead{If older Galactic Disk  member} 
$T_{eff}$ & $\sim 1410$ & K; see Section~\ref{sec-temp}\\
Age & $\le 1000$ & Myr \\
Mass & $<0.055$ & $M_\odot$\\
$\log g$ & $\sim 5.0$ & \\
\enddata
\end{deluxetable}

\begin{deluxetable}{lcl}
\tablewidth{0pc}
\tabletypesize{\footnotesize}
\tablenum{2} \label{tab2}
\tablecaption{2MASS J21481628+4003593}
\tablehead{
\colhead{Parameter} & 
\colhead{2M2148+40} & 
\colhead{Remarks}}
\startdata
Sp. Type & L6 FLD-G & Infrared,  \citet{2013ApJ...772...79A} \\
Sp. Type & L6 & Optical,  \citet{Looper:2008lr},\\
$\pi_{abs}$ & $124.07 \pm 0.55$ & mas \\
$\mu_{rel}$ & $901.9 \pm 0.3$ & mas \\
U & $-34.0 \pm 0.5$ & \kms \\
$m_{bol} $ & $15.12 \pm 0.03$ & Section~\ref{sec-temp}\\
$L/L_\odot$ & $(4.64 \pm 0.14) \times 10^{-5}$ & Section~\ref{sec-lum}\\
$M_J$ & $14.62 \pm 0.03$ & 2MASS \\
$M_H$ & $13.25 \pm 0.03$ & 2MASS \\
$M_{Ks}$ &$12.23 \pm 0.03$ & 2MASS \\
$M_{W1}$ &$11.21 \pm 0.02$ & WISE \\
$M_{W2}$& $10.70 \pm 0.02$ & WISE \\
$M_{W3}$& $10.13 \pm 0.06$ & WISE \\
\enddata
\end{deluxetable}

\end{document}